\documentclass[twocolumn,amssymb,preprintnumbers]{revtex4}

\usepackage{graphicx}
\usepackage{dcolumn}
\usepackage{bm}

\newcommand{\beq}{\begin{eqnarray}}
\newcommand{\beqa}{\begin{eqnarray*}}
\newcommand{\eeq}{\end{eqnarray}}
\newcommand{\eeqa}{\end{eqnarray*}}

\newcommand{\nn}{\nonumber}

\newcommand{\del}{\partial}
\newcommand{\ba}{\begin{array}}
\newcommand{\ea}{\end{array}}

\newcommand{\sch}{Schr{\" o}dinger }

\def\8{\infty}
\def\oh{\frac{1}{2}}

\def\d{\partial}

\def\undertext#1{\vtop{\hbox{#1}\kern 1pt \hrule}}

\def\VEV#1{\left\langle\,#1\,\right\rangle}

\def\dd#1{\frac{d}{d#1}}
\def\dbyd#1#2{\frac{d#1}{d#2}}

\def\pbyp#1#2{\frac{\partial#1}{\partial#2}}

\def\be{\begin{equation}}
\def\ee{\end{equation}}
\def\bea{\begin{eqnarray} & &}
\def\eea{\end{eqnarray}}

\def\rf#1{(\ref{#1})}

\def\t{\tilde}

\def\rfs#1{Eq.~\rf{#1}}

\begin{document}

\title{Spectra of pinned charge density waves with background current}

\author{V. Gurarie and J. Levinsen}
\affiliation{Physics Department, University of Colorado, Boulder, Colorado
  80309-0390}

\date{\today}

\begin{abstract}
We develop techniques which allow us to calculate the spectra of
pinned charge density waves with background current.
\end{abstract}

\maketitle


\section{Introduction}
The problem of a quantum mechanical particle whose motion is
governed by a random time-independent Hamiltonian has been
discussed in the literature for several decades. Because of its
relevance to the electrons' motion in disordered conductors, the
Hamiltonian which consists of a kinetic energy and a random scalar
potential is probably the most famous example of such a problem.
It is now well established that some of the wave functions of such
Hamiltonians are localized, the phenomenon usually referred to as
Anderson localization \cite{Anderson}. A wealth of new types of
behavior was discovered once random Hamiltonians constrained by
certain symmetries were studied. For example, the most general
random Hamiltonian would be complex hermitian. Constraining it to
be real (in other words, imposing time reversal invariance)
changes the localization properties of its wave functions. An
electron moving in a disordered conductor is described by a real
random Hamiltonian. Turning on a magnetic field makes the
Hamiltonian hermitian. This made the study of the crossover
between real and hermitian random Hamiltonians easily accessible
experimentally, and a subject of intense theoretical
investigations \cite{Efetov}.

Furthermore, it was discovered that a number of other symmetries
can also be imposed on random Hamiltonians, which  change their
behavior yet again. Altogether, there are ten symmetry classes of
random Hamiltonians, distinguished by nine different constraints
imposed on them \cite{AZ,Zirn}.

Yet other types of random Hamiltonians arise when they are
constrained not by a symmetry but by certain requirements their
spectra must satisfy. The most prominent example of that would be
systems with bosonic excitations \cite{vgjt2}. To be  specific,
consider an energy functional $E[\phi(x)]$, where $\phi(x)$ is
some function, which has the form
\begin{equation}
\label{eq:energy} E[\phi(x)] = \int_0^L dx~\left[ \oh \left( {d
\phi(x)\over dx} \right)^2 + h(\phi, x) \right].
\end{equation}
Here $h(\phi(x),x)$ is a random function whose form will be
specified below. Suppose $\phi_0(x)$ is a  minimum (local or
global) of this functional. We would like to study the normal
modes $\psi_n$ of oscillations around that minimum, described by
the equation
\begin{equation}
\label{eq:normal} -{d^2 \over dx^2} \psi_n(x) + \d^2_\phi h
(\phi_0(x),x)~\psi_n(x) = \epsilon_n \psi_n.
\end{equation}
\rfs{eq:normal} is a random Schr\"odinger-like equation. Yet it is
not equivalent to a particle moving in an arbitrary  random
potential. $\d^2_\phi h(\phi_0(x),x)$ is a random function of $x$,
but it is not an arbitrary random function. It is clear, for
example, that $\epsilon_n \ge 0$ for all $n$, which tells us that
$\d^2_\phi h(\phi_0(x),x)$, although random, has to be constrained
in such a way that all its eigenvalues are positive.

In most applications of \rfs{eq:energy}, $h(\phi,x)$ is chosen to
be a smooth function of $\phi$ and a rough function of $x$. For
example, $h(\phi,x)=A(x) \cos(\phi-\chi(x))$, where $A(x)$ and
$\chi(x)$ are random functions of $x$ uncorrelated at different
$x$, and $\chi(x)$ is uniformly distributed between $0$ and $2
\pi$. This leads to the two point correlation function
$$
\VEV{h(\phi,x)~h(\phi', x')} = \alpha ~\cos(\phi-\phi')~
\delta(x-x').
$$

Under this definition of $h(\phi,x)$, the problem described by
Eqs.~\rf{eq:energy} and \rf{eq:normal} has been extensively
studied in the context of pinned charge density waves
\cite{fukuyamalee}. Knowing the spectrum of \rfs{eq:normal}
allows, for example, to calculate the AC conductance of the charge
density wave.  It was first deduced in Refs.~\cite{AR} and
\cite{MF} that the density of states $\rho(\epsilon)$ of
\rfs{eq:normal} is given by $\rho(\epsilon)=\epsilon^{3 \over 2}$
if $\epsilon \ll \epsilon_c$ and if $\phi_0(x)$ is a global
minimum of the energy functional \rfs{eq:energy}, where
$\epsilon_c\sim \alpha^{2/3}$ is the crossover scale. If, on the
other hand, $\phi_0(x)$ is but a local minimum of \rfs{eq:energy},
then $\rho(\epsilon) \sim \epsilon$, $\epsilon \ll \epsilon_c$.

A more detailed approach to the problem specified by
\rfs{eq:normal} was developed in Ref.~\cite{vgjt}. It was shown in
that work that the potential $\d^2_\phi h(\phi_0(x),x)$ of the
``Schr\"odinger" equation \rfs{eq:normal} can always be
represented as
$$
\d^2_\phi h(\phi_0(x),x) = \dbyd{V(x)}{x}+V^2(x),
$$
where $V(x)$
is some new random function. As a result, \rfs{eq:normal} is
equivalent to \begin{equation} \label{eq:chir} {\cal H} \left(
\matrix {\psi_n(x) \cr \phi_n(x)} \right) = \omega_n \left(
\matrix {\psi_n(x) \cr \phi_n(x)} \right),
\end{equation}
where
\begin{equation}
\label{eq:calH} {\cal H} = \left( \matrix{ 0 & \dd{x} + V(x) \cr
-\dd{x} + V(x) & 0 } \right),
\end{equation}
and $\omega_n^2=\epsilon_n$. Now ${\cal H}$ is an example of a
Hamiltonian constrained by a symmetry and can be solved by
techniques developed in that context. The symmetry of ${\cal H}$
is usually referred to as chiral symmetry. It is expressed by the
relation $\sigma_3 {\cal H} \sigma_3 = - {\cal H}$, which holds
true for any random $V(x)$ \cite{AZ}. Here $\sigma_3$ is the usual
Pauli matrix.

The full solution of \rfs{eq:normal}, with the help of the mapping
to \rfs{eq:chir}, demonstrated that there indeed exists the
crossover energy scale $\epsilon_c$. At $\epsilon_n \ll
\epsilon_c$ all the eigenfunctions $\psi_n(x)$ are localized with
the localization length $\xi_n \sim \epsilon_c^{-{1 \over 2}}$,
which is independent of $n$. At $\epsilon_n \gg \epsilon_c$, the
eigenfunctions are still localized, but with the localization
length which increases with $\epsilon_n$ as $\xi_n \sim
\epsilon_n$. The density of states at  $\epsilon_n \gg \epsilon_c$
is given by $\rho(\epsilon) \sim \epsilon^{-{1 \over 2}}$.
Finally, the density of states at  $\epsilon_n \ll \epsilon_c$ is
given by $\rho(\epsilon) \sim \epsilon^{3 \over 2}$ if $\phi_0(x)$
is a global minimum of \rfs{eq:energy} and by $\rho(\epsilon) \sim
\epsilon$ if $\phi_0(x)$ is a local minimum of \rfs{eq:energy}.

In this paper we would like to show that the same techniques which
proved useful in solving \rfs{eq:normal} can also be used  to
solve another related problem, which we formulate below. Consider
an equation
\begin{equation}
\label{eq:flux} -j {d^2 \phi \over dx^2} - \gamma \dbyd{\phi}{x} +
\d_\phi h(\phi,x)=0,
\end{equation}
where $j$ and $\gamma$ are some parameters. If $\gamma=0$, then
this equation is equivalent to the minimization condition of the
energy \rfs{eq:energy}. If, on the other hand, $\gamma>0$, then
this defines a new problem. \rfs{eq:flux} was suggested in
Ref.~\cite{lrjt} to describe pinned charge density waves with
background current. Normal modes of oscillations of such pinned
charge density wave are given by
\begin{equation}
\label{eq:norm} \left[-j {d^2 \over dx^2}  -\gamma \dd{x}
+\d^2_\phi h (\phi_0(x),x) \right] \psi_n(x) = \epsilon_n
~\psi_n(x),
\end{equation}
where $\phi_0(x)$ is a solution to \rfs{eq:flux}. In this paper we
are going to present the solution to the problem defined by
\rfs{eq:flux} and \rfs{eq:norm}.

\begin{figure}[bt]
\includegraphics[height=2in]{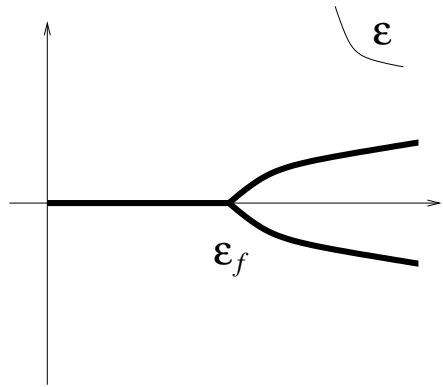}
\caption{\label{curve} The eigenvalues $\epsilon_n$ of the
\rfs{eq:norm} lie on this curve in the complex plane.}
\end{figure}

The operator in the square brackets of \rfs{eq:norm} is
nonhermitian. As a result, the eigenvalues $\epsilon_n$ do not
have to be real. It is well known, however, that the eigenvalues
of equations of the type \rfs{eq:norm} always lie along one
dimensional curves in the complex plane \cite{gk}. In this paper,
we show that these curves take the shape depicted on Fig.
\ref{curve}. The fork point $\epsilon_f$ cannot be found exactly.
However, it is still possible to define the crossover scale
$\epsilon_c$. All the states with energy less than $\epsilon_c$
have the same localization length $\xi$. As the energy is
increased past $\epsilon_c$, the localization starts to grow and
eventually diverges at $\epsilon_f$ (which is always bigger than
$\epsilon_c$). The states corresponding to complex values of
energy are delocalized.

We introduce the technique which allows to compute $\epsilon_c$
and $\xi$. We calculate them in the regime where $\gamma /(j
\alpha)^{1/3}\gg 1$ and find them to be $\epsilon_c = {\gamma^2/4
j}$ and $\xi = 2 \gamma^2/\alpha$. We also calculate the density
of states along the part of the curve depicted on Fig.~\ref{curve}
which lies on the real axis and find it to be $\rho(\epsilon) \sim
\epsilon$.

The rest of this paper is organized as follows. In section
\ref{sec:bur} we map the problem defined by \rfs{eq:norm} to a
random chiral Hamiltonian. In section \ref{sec:loc} the
localization length of normal mode oscillations will be derived in
the limit of large background current and in section \ref{sec:lar}
this result will be compared with the usual Larkin length of the
problem. Section \ref{sec:den} contains the calculation of the
density of the low lying states. Finally, in Section \ref{sec:cur}
we discuss the response of a pinned charge density wave with
background current to an electric field using the formalism
developed here.


\section{Mapping to a Chiral Hamiltonian\label{sec:bur}}

The first step of the solution is to map \rfs{eq:norm} to a more
tractable form of a random chiral Hamiltonian. Following the
general guidelines of Ref.~\cite{vgjt}, we treat \rfs{eq:norm} as
the equation of motion of a particle with coordinate $\phi$,
moving in time $x$, in the presence of viscous forces given by the
$\gamma$ term. We then pass from the Lagrangian to Eulerian
description of the motion. Introducing the velocity function
$u(\phi_0(x))\equiv\del_x\phi_0(x)$ the equation of motion
\rfs{eq:norm} becomes \beq j\del_xu+j u \del_\phi u +\gamma u & =
& \del_\phi h(\phi_0(x),x). \label{burgers} \eeq The equation thus
obtained is similar to the equation of motion of a one dimensional
fluid without pressure, usually referred to as Burgers equation.
However, it also features the $\gamma u$ term, which can be
interpreted as a  kind of viscous friction. The Burgers equation
with this term is novel and has not been considered in the
literature before.

Performing a derivative on \rfs{burgers} with respect to $\phi$
results in \beq j\del_x\del_\phi u+j(\del_\phi u)^2+ju\del^2_\phi
u+\gamma\del_\phi u & = & \del^2_\phi h. \label{funcder} \eeq We
define the gradient of the velocity along the solution to
\rfs{eq:flux}, $\phi_0$, as $F(x) \equiv\del_\phi u(\phi_0(x),x)$.
$F(x)$ can be  related to $h(\phi,x)$ with the help of
Eq.~(\ref{funcder}) \beq \label{vdif} j\dbyd{F}{x}+jF^2 +\gamma F&
= & \del^2_\phi h(\phi_0(x),x). \label{lang1} \eeq Thus
Eq.~(\ref{eq:norm}) becomes \beq \left[-{d^2 \over
dx^2}-\frac\gamma j\dd{x} +\dbyd{F}{x}+F^2+\frac\gamma j
  F\right]\psi_n & = &
\frac{\epsilon_n}{j} \psi_n, \eeq or in a more symmetric form \beq
\left[\frac {d}{d x}+F+\frac\gamma j \right]\left[-\frac {d}{d
    x}+F\right] \psi_n
& = & \frac{\epsilon_n}{j}\psi_n. \label{almostsymform} \eeq The
operator on the left-hand-side of Eq.~(\ref{almostsymform}) is not
Hermitian. This may be remedied by using a trick due to Hatano and
Nelson \cite{hn}. Writing \beq \psi_n(x) & = & e^{-\frac\gamma{2j}
x}\tilde{\psi}_n(x) \label{hatanonelson} \eeq the equation becomes
\beq \left[\dd{x}+F+\frac\gamma{2j} \right]\left[-
\dd{x}+F+\frac\gamma{2j}\right] \tilde{\psi}_n & = & \frac{
\epsilon_n}{j}\tilde{\psi}_n. \label{herm} \eeq Thus we mapped our
equation into an eigenvalue problem for a Hermitian operator. This
operator can also be rewritten in the form similar to
\rfs{eq:chir},
$$
{\cal H} \left( \matrix {\t \psi_n(x) \cr \t \phi_n(x)} \right) =
{\omega_n \over \sqrt{j}} \left( \matrix {\t \psi_n(x) \cr \t
\phi_n(x)} \right),
$$
where ${\cal H}$ is still given by \rfs{eq:calH},
$\omega_n^2=\epsilon_n$, and $$V(x)=F(x)+{\gamma \over 2j}.$$
However, as argued by Hatano and Nelson, the transformation
\rfs{hatanonelson} is valid only as long as $\t \psi_n$ decays
asymptotically as $\exp(-|x|/{\t \xi})$, with the localization
length ${\t \xi} < 2 j /\gamma$. For $\t \psi_n$ whose
localization length obeys this condition, \rfs{almostsymform} is
equivalent to \rfs{herm}, and therefore, $\epsilon_n$ is real and
positive. For other eigenfunctions $\t \psi_n$ whose localization
length is larger than $2 j/\gamma$, \rfs{herm} is no longer
equivalent to \rfs{almostsymform}.

Notice that the localization length $\t \xi_n$ of $\t \psi_n$ is
related to the localization length $\xi_n$ of $\psi_n$ as in
\begin{equation}
\label{eq:locrel} \xi_n = \left( {1 \over \t \xi_n} - {\gamma
\over 2j} \right)^{-1},
\end{equation}
as long as $\t \xi < 2j/\gamma$.



\section{The Localization length \label{sec:loc}}


Random chiral Schr\"odinger equations of the form (\ref{herm})
have been investigated by A. Comtet, J. Debois, and C. Monthus
\cite{comtet}. It was determined that the important parameter for
these equations is $\omega_c = \sqrt{j} \VEV{V(x)}$, and
correspondingly $\epsilon_c \equiv\omega_c^2$. For all
wavefunctions whose energy $\omega_n<\omega_c$, the localization
length is constant and is given by $\t \xi=1/\VEV{V(x)}$. For
$\omega_n>\omega_c$, the localization length quickly increases,
being asymptotically proportional to $\omega_n^2$.

The crucial test for our theory is, therefore, whether
$\VEV{V(x)}$ is bigger or smaller than $\gamma/2j$. As we will see
below, $\VEV{V(x)}>\gamma/2j$. As a result, \rfs{herm} is
equivalent to \rfs{almostsymform} and consequently, to
\rfs{eq:norm}, at $\epsilon<\epsilon_c$. Thus at
$\epsilon<\epsilon_c$ all the eigenvalues of \rfs{eq:norm} are
real and positive. On the other hand, there exist some
$\epsilon_f>\epsilon_c$ where the localization length of
$\t\psi_n$ becomes equal to $2j / \gamma$. At that point, the wave
functions $\psi_n$ become delocalized, in accordance with
\rfs{eq:locrel}. At ${\rm Re}~\epsilon>\epsilon_f$, the
eigenvalues of \rfs{eq:norm} are no longer real and come in
complex conjugate pairs. This justifies the picture presented on
Fig.~\ref{curve}.

In what follows, we proceed to calculate $\epsilon_c$. In terms of
$V(x)$, \rfs{lang1} becomes \beq
 \dbyd{V}{x}+V^2-\frac{\gamma^2}{4j^2} & = &
\frac{\del_\phi^2 h(\phi_0(x),x)}j. \label{lang2} \eeq This
equation has the form of a Langevin equation with random force
given by $\del_\phi^2h$, whose correlator is given by (see
Ref.~\cite{vgjt2} for justification of $\del_\phi^2h$ as a random
white noise)
$$\left<\del_\phi^2h(\phi_0(x),x)~\del_\phi^2h(\phi_0(y),y)\right>=\alpha~\delta(x-y).$$

In accordance with the theory of Langevin equations, if $$
\dbyd{V}{x}+g(V)  =  f(x) $$ with $f(x)$ being uncorrelated at
different values of $x$,
$\left<f(x)f(x')\right>=\alpha\delta(x-x')$, then  the probability
$P(v,x)=\left<\delta(v-V(x)\right>$ of observing $V(x)=v$ at the
position $x$ obeys the Fokker-Planck equation $$ \frac {dP}{dt}  =
\frac\del{\del v}\left[\frac\alpha 2\frac\del{\del
    v}+g(v)\right]P(v,x).
$$

In the present case, we are interested in the probability of
observing a particular value for $V(x)$ together with the
probability that a particular solution $\phi_0(x)$ of
\rfs{eq:flux} is chosen. The relevant quantity describing this
joint probability is $${\cal P}(v,x)=\VEV{{\delta(v-V(x)) \over
\rho(x)}},$$ where $\rho(x)$ is the density of solutions
$\phi_0(x)$ which in turn obeys the continuity equation
$$
\dbyd{\rho}{x}+\rho V =0.
$$
As a result, we find the Fokker-Planck equation  $$ \frac{d{\cal
P}}{dx}  =  \frac{\del}{\del
v}\left[\frac\alpha{2j^2}\frac{\del}{\del
    v} + v^2 -\frac{\gamma^2}{4j^2}\right]{\cal P} + \lambda v {\cal
    P}.
$$ Here $\lambda=1$, however, we will keep the more general
notation of $\lambda$ for convenience at a later point in the
calculations.

A common approach to the Fokker-Planch equations is to map them
into the Schr{\" o}dinger equation with the help of
$${\cal P} = \exp\left[{-{v^3 j^2\over 3 \alpha} + {v \gamma^2 \over 4 \alpha}}\right]
 \Psi.$$
This gives \be -\dbyd{\Psi}{x}= \left[
-\frac\alpha{2j^2}\frac{\del^2}{\del
  v^2}+U(v) \right]\Psi   \label{eq:schr}. \ee
Here
\begin{equation}
\label{eq:potential}
U(v)=\left[\frac{j^2}{2\alpha}v^4-\frac{\gamma^2}{4\alpha}v^2-v
  (1+ \lambda)
+ \frac{\gamma^4}{32\alpha j^2}\right].
\end{equation}

The Feynman path integral formulation of the Schr\"odinger
equation \rfs{eq:schr} allows us to find the average of $V$ using
\beq \ell \left<V\right> & = & \frac{\int {\cal D}V(x)~
e^{-S[V]}\int_0^\ell dx V(x)}{\int {\cal D}V(x)
  e^{-S[V]}} \nn \\
& = & \left. \frac d{d\lambda}\log\int {\cal D}V(x) e^{-S[V]}
\right|_{\lambda=1}, \label{eq:path}\eeq for $\ell \rightarrow
\infty$. Here $S[V]$ is the imaginary time action which
corresponds to the quantum mechanics \rfs{eq:schr},
\begin{equation}
\label{eq:action} S[V]= \int_0^\ell dx~\left[ {j^2 \over 2 \alpha}
\left(\dbyd{V}{x} \right)^2 + U(V) \right].
\end{equation}

At large $\ell$ the path integral \rfs{eq:path} is dominated by
its ground state energy.  The average $V$ may now be calculated as
\beq \left<V\right> & = &
-\left.\frac{dE_0}{d\lambda}\right|_{\lambda=1}, \eeq where $E_0$
is the quantum mechanical ground state energy for a particle whose
classical action is given by \rfs{eq:action} and whose
Schr\"odinger equation reads $$ \left[ -{\alpha \over 2 j^2} {d^2
\over dv^2} + U(v) \right] \Psi = E \Psi
$$


If $\gamma=0$, then the ground state energy, together with its
derivative with respect to $\lambda$, can be estimated simply from
dimensional analysis as $\alpha^{1/3}/j^{2/3}$, to give $\omega_c
\sim \alpha^{1/3}/j^{1/6}$. This agrees with Ref.~\cite{vgjt}. If
$\gamma>0$, then dimensional analysis is of no help, since
$\omega_c$ can now depend on the dimensionless ratio $\gamma /(j
\alpha)^{1/3}$. The only way to find $\omega_c$ and thus the
localization length of the low lying states is by finding the
ground state energy $E_0$.

By a suitable rescaling $v = y \alpha^{1/3}/j^{2/3}$, $\t E = E
j^{2 / 3} / \alpha^{1 / 3}$, we can bring the Schr\"odinger
equation to the form
\begin{equation}
\label{eq:schdiml} \left[- \oh {d^2 \over dy^2}+\oh \left(y^2-c^2
\right)^2 - \left(1 + \lambda \right) y \right] \Psi  = \t E \Psi,
\end{equation} where
$$c={\gamma \over 2 \left( \alpha j \right)^{1 \over 3}}.$$

In general, it is not possible to find the ground state energy of
this Schr\"odinger equation exactly. We are going to find it only
in the limit when $c \gg 1$. The potential in \rfs{eq:schdiml} has
two minima. The minimum at positive $x$ is the global minimum and
is located at \begin{equation} y_{\min} = c+{(1+\lambda) \over 4
c^2} + {\cal O}(c^{-5}). \label{vmin} \end{equation} The simplest
approximation to the ground state energy is to set it equal to the
value of the potential at the minimum. This is given by $$ \t
E_{0,0} =  -(1+\lambda)c -{(1+\lambda)^2 \over 8 c^2} +{\cal
O}(c^{-5}). $$ To add quantum fluctuations to the problem, we have
to consider the \sch equation (\ref{eq:schdiml}) evaluated around
the minimum \rfs{vmin}. Approximating the potential of
\rfs{eq:schdiml} by a quadratic potential around the point
$x_{\min}$ we find the oscillator ground state energy to be $$ \t
E_{0,qf}  =  c+{3 (1+\lambda) \over 8 c^2}+{\cal
  O}(c^{-5}).
$$
In principle, the cubic and quartic term in the expansion of the
potential around the minimum also contribute to the ground state
energy, but these only have dependence upon $\lambda$ at the
${\cal O}(c^{-5})$ level and thus do not contribute to
$\left<V\right>$ at lower orders.

Gathering the terms, it is seen that as a function of $\lambda$
\beq \left<V\right> & = & -\frac{dE_0}{d\lambda} = -\dd{\lambda}
 \left( \t E_{0,0} + \t E_{0,qf} \right){\alpha^{1 \over 3} \over
j^{2 \over 3}} \nn \\
& = &
\frac{\gamma}{2j}+\alpha\left(\lambda-\oh\right)\frac1{\gamma^2}+{\cal
O}(\gamma^{-5}). \nn \eeq Finally, substituting $\lambda=1$, we
find
$$
\VEV{V}=\frac{\gamma}{2j}+{\alpha\over 2 \gamma^2}+{\cal
O}(\gamma^{-5}).
$$
Therefore, at large $\gamma$ the energy scale $\omega_c$ is given
by
\begin{equation} \label{eq:wc}
\omega_c={\gamma \over 2 \sqrt{j}}. \end{equation} We see that
$\VEV{V}>\gamma/2j$, and the localization length of the low lying
hermitian states $\t \psi_n$ is consequently smaller than $2
j/\gamma$. This allows us to deduce that \rfs{eq:norm} is indeed
equivalent to \rfs{herm} at $\epsilon_n<\epsilon_c$. The
localization length of $\psi_n$ at $\epsilon_n<\epsilon_c$ can be
deduced from \rfs{eq:locrel} to be
\begin{equation}
\label{eq:local} \xi= \left(\VEV{V}-{\gamma \over 2 j}
\right)^{-1} = {2 \gamma^2 \over \alpha} + {\cal O}(\gamma^{-1}).
\end{equation}

\begin{figure}[bt]
\includegraphics[height=1.6in]{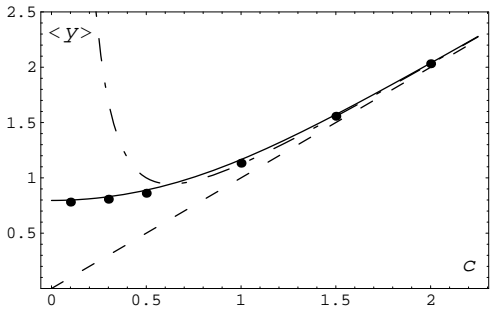}
\caption{$\VEV{y}$ in the ground state of \rfs{eq:schdiml} as a
function of $c$. The dots represent $\VEV{y}$ evaluated
numerically at various values of $c$. The solid line is $\VEV{y}$
calculated in the harmonic approximation to the potential in
\rfs{eq:schdiml}. The dash-dotted line represents the first two
terms of the perturbative calculation $c+{1 \over 8 c^2}$.
Finally, the dashed line is a straight line of the slope 1, which
demonstrates that $\VEV{y}>c$.} \label{localization}
\end{figure}

The calculation presented here justifies Fig.~\ref{curve} at large
$\gamma$. In order to see that the picture of low lying localized
states persist at all values of $\gamma$ we need to show that
$\VEV{V}$ is always greater than $\gamma/2j$. This amounts to
showing that in the Schr\"odinger equation \rfs{eq:schdiml} the
following relation  always holds true \begin{equation}
\label{eq:theor} \VEV{y}\equiv \int dy~ |\psi_0(y)|^2 y=-\left.
\dbyd{\t E_0}{\lambda}\right|_{\lambda=1}
> c.
\end{equation} Here $\psi_0(x)$ is the ground state wave function
of \rfs{eq:schdiml}, at $\lambda=1$. We do not know how to show
this analytically. We investigated \rfs{eq:schdiml} numerically.
Fig. \ref{localization} shows the values of $-d\t
E_0/d\lambda=\VEV{y}$, at $\lambda=1$, plotted versus $c$. We see
that \rfs{eq:theor} does seem to hold.

\section{The Larkin length \label{sec:lar}}
In the analysis of \rfs{eq:flux} it is customary to introduce the
notion of Larkin length $L$. $L$ is defined as size of a box in
which the average $\VEV{\phi^2}$ becomes of the order of 1. Thus
at distances bigger than $L$, $\phi_0(x)$ becomes rough, while at
shorter distances it can be thought of as smooth. We are now going
to see that in our problem the localization length of low lying
states of \rfs{eq:norm} is of the order of Larkin length.

To calculate the Larkin length, we follow the procedure described
in Ref.~\cite{lrjt}. We consider the equation \rfs{eq:flux}. Let
us assume that $\phi$ is small and neglect the $\phi$ dependence
of $h$. In Fourier space this equation of motion is $$ (i\gamma k
+ j k^2)\phi_k  =  -\del_\phi h(0). $$ Thus \beq
\left<\left|\phi_k\right|^2\right> & = & \frac{\left<\d_\phi h(0)
~\d_\phi h(0)\right>}{\gamma^2 k^2+j^2 k^4}.\nn \eeq Now the
Larkin length is defined as the value of $L$ for which \beq
1 & \sim & \left<\phi^2_0(x)\right> \nn \\
& = & \int_{L^{-1}}^\infty {dk \over 2\pi} \frac\alpha{\gamma^2 k^2+j^2k^4} \nn \\
& \sim & \frac{L\alpha}{\gamma^2}-
\frac{j\alpha}{\gamma^3}\tan^{-1}(L\gamma/j). \nn \eeq From the
asymptotic form of $\tan^{-1}$ in the large-$\gamma$ limit it is
found that \beq L & \sim &
\frac{\gamma^2}\alpha+\frac\pi2j\gamma^{-1}-j^2\alpha\gamma^{-4}+{\cal
  O}(\gamma^{-7}). \nn
\eeq At leading order $L$ turns out to be proportional  to the
localization length obtained above \rfs{eq:local}. This coincides
with the behavior of the pinned charge density waves without the
background current \cite{vgjt}. However, the localization length
and the Larkin length have different functional dependence on
$\gamma$ in lower orders.


\section{Density of states \label{sec:den}}
In this section we calculate the density of low lying states for
our problem. Due to the form of the potential \rfs{eq:potential}
which does not contain a cubic term, the density of low lying
states may be easily calculated using the methods of ref.
\cite{vgjt2}. This is not immediately obvious, so in this section
we will repeat the argument.

At energies below an energy of the order of $\left< V\right>$ the
energy eigenvalues are real. Above, the eigenvalues trace out a
one-dimensional spectrum in the complex plane, Fig.~\ref{curve}.
Let us concentrate on low lying states $\epsilon_n<\epsilon_c$,
which are all real. To calculate their density of states
$\rho(\epsilon)$ we use the equivalence of \rfs{eq:norm} and
\rfs{herm} in this regime.

The integrated density of low-lying states can be found using the
discussion in Ref.~\cite{vgjt2} as \beq N(\omega) & = &
\left<\delta \left( \int_{x_1}^{x_2}dx~V(x)+a \right)\right>, \nn \\
 a&=&-\log(\omega), \nn \eeq with the average being over realizations of
$V(x)$.  The density of states is then obtained from
$\rho(\omega)=dN/d\omega$.

Through a standard representation of the $\delta$-function \beq
N(\omega) & = & \left<\int_{-\infty}^\infty
\frac{d\alpha}{2\pi}e^{i\alpha\left( \int_{x_1}^{x_2}dx
V(x)+a\right)}\right>, \nn \eeq and defining $S(\alpha)$ by \beq
\exp(-\ell S(\alpha)) & = & \left<\exp\left(i\alpha
\int_{x_1}^{x_2}dx V(x)\right)\right>, \label{sofa} \eeq with
$\ell=x_2-x_1$ the integrated density of states is \beq N(\omega)
& = & \int_{-\infty}^\infty\frac{d\alpha}{2\pi}e^{i\alpha a-\ell
  S(\alpha)}. \nn
\eeq At large $\ell$, $S(\alpha)$ is independent of $\ell$ as will
be argued below. We assume in the following that  a certain value
of $\ell$ maximizes the probability of observing a fluctuation of
size $-\log\omega$. This is justified, since due to the form of
the potential $U(V)$, \rfs{eq:potential}, it is not very likely to
observe a very large negative fluctuation, and it is more likely
to see a smaller negative fluctuation, which exists over a longer
interval of $x$.  We now approximate $N(\omega)$ by its value at
the saddle point, determined from \beq ia-\ell\frac{\del
S}{\del\alpha} & = & 0, \nn \eeq where $\alpha=\alpha_0(\ell)$ at
the saddle point. Thus $N(\omega)\sim\exp(ia\alpha_0-\ell
S(\alpha_0(\ell)))$ and the maximization with respect to $\ell$
results in \beq 0 & = &
ia\frac{\del\alpha_0}{\del\ell}-S(\alpha_0(\ell)) -\ell\frac{\del
  S}{\del\alpha}\frac{\del\alpha}{\del\ell} \nn \\
& = & -S(\alpha_0(\ell)). \nn \eeq Defining $\beta_0\equiv
-i\alpha_0$ it is seen that $N(\omega)\sim \omega^\beta_0$ where
$\beta_0$ should be chosen such that $S(\beta_0)$ vanishes.

Making use of our results from section \ref{sec:loc} where the
\sch equation Eq.~(\ref{eq:schr}) was found with the potential
taking
 the form Eq.~(\ref{eq:potential}), Eq.~(\ref{sofa}) becomes
\beq & & \exp(-\ell S(\beta_0)) \nn \\ & = & \frac{\int {\cal
    D}V(x)
\exp\left( -S[V]\right)\exp\left(-\beta_0\int_{x_1}^{x_2}dx
V\right)} {\int {\cal D}V(x)\exp\left(-S[V] \right)  }
\label{spath}, \eeq
with $S[V]$ given by \rfs{eq:action}. At large lengths $\ell$ the
path integrals are dominated by the ground state energy, i.e. \beq
\exp(-\ell S(\beta_0)) & = & \frac{\exp(-\ell
E_0\left\{-1-\lambda+\beta_0 \right\})}{\exp(-\ell
E_0\left\{-1-\lambda\right\})}, \nn \eeq where the notation
$E_0\left\{s\right\}$ represents the ground state of the
Schr\"odinger equation \rfs{eq:schr} with the potential
\rfs{eq:potential} where $s$ is substituted in place of
$1+\lambda$. Furthermore, since all terms except the linear are
invariant under $V(x)\to -V(x)$, we see that
$E_0\left\{-1-\lambda+\beta_0\right\}=
E_0\left\{1+\lambda-\beta_0\right\}$. Thus $\beta_0=2+\lambda$
obviously solves the requirement.  Note that the above argument
hinged on the fact that the potential $U(V)$ does not contain a
cubic term. We conclude that with $\lambda=1$ \beq N(\omega) &
\sim & \omega^4  = \epsilon^2, \nn \eeq and \beq \rho(\omega) &
\sim & \omega^3, \ \rho(\epsilon)=\dbyd{N(\epsilon)}{\epsilon}
\sim \epsilon . \label{eq:DoS} \eeq

We would like to compare this with the result obtained for the
pinned charge density wave problem with $\gamma=0$. There
$\rho(\omega) \sim \omega^4$ if $\phi_0$ is a global minimum of
the energy functional \rfs{eq:energy} and $\rho(\omega) \sim
\omega^3$ if the minimum is local. At $\gamma>0$, however,
\rfs{eq:flux} is no longer a minimization condition of any
functional and the notion of global minimum no longer exist.

\section{Driven Pinned Charge Density Waves with
background Current \label{sec:cur}} Consider a pinned charge
density wave driven by an electric field at frequency $\omega_0$
(see e.g. Ref.~\cite{Natt} for a review),
$$ \nu {d \phi \over dt}-j {d^2 \phi \over dx^2} - \gamma \dbyd{\phi}{x} + \d_\phi
h(\phi,x)=E(x) \cos (\omega_0 t ).
$$
To find the linear response to a small electric field, we write
$\phi=\phi_0(x)+\psi(x,t)$, where $\phi_0(x)$ satisfies
\rfs{eq:flux}, and find
$$
\left[\nu {d \over dt}-j {d^2 \over dx^2}  -\gamma \dd{x}
+\d^2_\phi h (\phi_0(x),x) \right] \psi(x,t) = E(x) e^{i \omega_0
t}.
$$
The solution to this equation can now easily be found
$$
\phi(x,t)=\phi_0(x)+  e^{i \omega_0 t}~\sum_n \int dy~E(y)
{\psi_n(x) \psi_n(y) \over i \nu\omega_0 + \epsilon_n },
$$
where $\psi_n$ and $\epsilon_n$ are defined in \rfs{eq:norm}.
The current carried by the charge density wave is $I \sim
\pbyp{\psi}{t}$, and is thus given by
$$ I(x) \sim i \omega_0
e^{i \omega_0 t}~\sum_n \int dy~E(y) {\psi_n(x) \psi_n(y) \over i
\nu\omega_0 + \epsilon_n }.
$$
The properties of $\psi_n$, $\epsilon_n$ found in this paper help
to determine the response of the charge density wave to a small
external electric field. For example, if the electric field $E(x)$
is uniform in space, we can use the results of this paper which
show that at $\epsilon_n < \epsilon_c$, the wave functions
$\psi_n(x)$ are all localized with the same localization length
$\xi$. Therefore, $\int dx~\psi_n(x) \propto \sqrt{\xi}$ and is
independent of $n$, as long as $\epsilon_n<\epsilon_c$. On the
other hand, the wave functions which correspond to large ${\rm
Re}~\epsilon_n$, are delocalized and oscillate fast, so that $\int
dx~\psi_n(x)$ goes to zero with increasing $n$ quickly. Therefore,
the current reduces to \beq I & \sim & i \omega_0 e^{i \omega_0 t}
\int_0^{\epsilon_c} d \epsilon~
\rho(\epsilon) {1 \over i \nu\omega_0 + \epsilon} \nn \\
& \sim & i \omega_0 e^{i\omega_0
t}\left[\epsilon_c-i\nu\omega_0\log{\epsilon_c+i\nu\omega_0 \over
i\nu\omega_0}\right], \eeq where $\rho(\epsilon) \sim \epsilon$,
as found in this paper, \rfs{eq:DoS}.




\section{Conclusions \label{sec:con}}
We have demonstrated how the problem of normal mode oscillations
of the one dimensional pinned charge density waves with background
current may be solved by mapping into chiral random Hamiltonians.
Using the methods of Comtet et al \cite{comtet} we have determined
the localization length $\xi$ of the  low lying normal modes. This
localization length turns out to be proportional to the Larkin
length of this system. The density of the lowest lying states has
also been obtained, giving the power law
$\rho(\omega)\sim\omega^3$. This result differs from the result
$\rho(\omega)\sim\omega^4$ of Refs. \cite{AR,vgjt} due to the fact
that turning on background current removes the notion of a global
ground state.

{\bf Acknowledgments} The authors wish to thank Leo Radzihovsky for useful
discussions. J.L. also wishes to acknowledge support from the Danish Research
Agency and the Danish-American Fulbright Commission.


\end{document}